\documentclass[%
 reprint,
superscriptaddress,
 amsmath,amssymb,
 aps,
prb,
]{revtex4-2}

\usepackage{graphicx}% Include figure files
\usepackage{dcolumn}% Align table columns on decimal point
\usepackage{bm}% bold math
\begin{document}

\preprint{APS/123-QED}

\title{Magnetic field-driven transition between valence bond solid and antiferromagnetic order in distorted triangular lattice}
% Force line breaks with \\

\author{Yasuhiro Shimizu}
\affiliation{Department of Physics, Nagoya University, Chikusa, Nagoya 464-8602, Japan.}
\author{Mitsuhiko Maesato}
\affiliation{Division of Chemistry, Graduate School of Science, Kyoto University, Sakyo-ku, Kyoto 606-8502, Japan.}
\author{Makoto Yoshida} 
\affiliation{Institute for Solid State Physics, University of Tokyo, Kashiwa, Chiba 277-8581, Japan.}
\author{Masashi Takigawa}
\affiliation{Institute for Solid State Physics, University of Tokyo, Kashiwa, Chiba 277-8581, Japan.}
\author{Masayuki Itoh}
\affiliation{Department of Physics, Nagoya University, Chikusa, Nagoya 464-8602, Japan.}
\author{Akihiro Otsuka}
\affiliation{Division of Chemistry, Graduate School of Science, Kyoto University, Sakyo-ku, Kyoto 606-8502, Japan.}
\affiliation{Research Center for Low Temperature and Materials Sciences, Kyoto University, Sakyo-ku, Kyoto 606-8501, Japan}
\author{Hideki Yamochi}
\affiliation{Division of Chemistry, Graduate School of Science, Kyoto University, Sakyo-ku, Kyoto 606-8502, Japan.}
\affiliation{Research Center for Low Temperature and Materials Sciences, Kyoto University, Sakyo-ku, Kyoto 606-8501, Japan}
\author{Yukihiro Yoshida}
\affiliation{Division of Chemistry, Graduate School of Science, Kyoto University, Sakyo-ku, Kyoto 606-8502, Japan.}
\affiliation{Faculty of Agriculture, Meijo University, Tempaku-ku, Nagoya 468-8502, Japan.}
\author{Genta Kawaguchi}
\affiliation{Division of Chemistry, Graduate School of Science, Kyoto University, Sakyo-ku, Kyoto 606-8502, Japan.}
\author{David Graf}
\affiliation{National High Magnetic Field Laboratory, Florida State University, Tallahassee, 32310 FL.}
\author{Gunzi Saito}
\affiliation{Faculty of Agriculture, Meijo University, Tempaku-ku, Nagoya 468-8502, Japan.}
\affiliation{Toyota Physical and Chemical Research Institute, Nagakute, Aichi 480-1192, Japan.}

\date{\today}% It is always \today, today,

\begin{abstract}
A molecular Mott insulator $\kappa$-(ET)$_2$B(CN)$_4$ [ET = bis(ethylenedithio)tetrathiafulvalene] with a distorted triangular lattice exhibits a quantum disordered state with gapped spin excitation in the ground state. $^{13}$C nuclear magnetic resonance, magnetization, and magnetic torque measurements reveal that magnetic field suppresses valence bond order and induces long-range magnetic order above a critical field $\sim 8$ T. The nuclear spin-lattice relaxation rate $1/T_1$ shows persistent evolution of antiferromagnetic correlation above the transition temperature, highlighting a quantum spin liquid state with fractional excitations. The field-induced transition as observed in the spin-Peierls phase suggests that the valence bond order transition is driven through renormalized one-dimensionality and spin-lattice coupling. 
\end{abstract}

\maketitle

%\tableofcontents
\section{Introduction}
	Quantum magnets featured by macroscopic quantum entanglements host exotic quasiparticles and topological order \cite{Anderson, Wen, Senthil}. In frustrated spin systems, a valence bond order state with translational symmetry breaking competes with the magnetic order and quantum spin liquid states \cite{Read, Alicea, Hauke,Tocchio}. The phase transition between magnetic order and valence bond crystal phases may occur through continuous quantum phase transition \cite{Senthil, Sachdev}, instead of the first-order transition that separates the two phases having competing order parameters in Landau-Ginzburg-Wilson paradigm. The realization of deconfined quantum criticality with fractional excitation has been a fundamental issue of many-body quantum physics \cite{Read, Senthil2, Jian, Scammell}.

Under an intense magnetic field, the valence bond order state is destabilized and exhibits a transition into long-range order states \cite{Zapf, Giamarchi}. The quasiparticle in the spin-dimer state is a spin $S = 1$ triplon localized on the lattice site. The field-induced transition involving long-range order of transverse magnetization is regarded as Bose-Einstein condensation of triplons, as observed in several quantum magnets \cite{Shirasawa, Yamada, Vyaselev, Mazurenko, Sebastian, Kramer, Lancaster, Zapf, Giamarchi}. On one-dimensional (1D) systems, the spin-Peierls transition occurs through spin-lattice coupling and is suppressed by forming a soliton lattice under high magnetic field \cite{Hayashi, Horvatic, Casola, Kiryukhin}. By contrast the spin-singlet formation in 2D frustrated spin systems can be highly degenerate and involve fractionalized quasiparticles \cite{Senthil2}. 

\begin{figure}
	\includegraphics[scale=0.55]{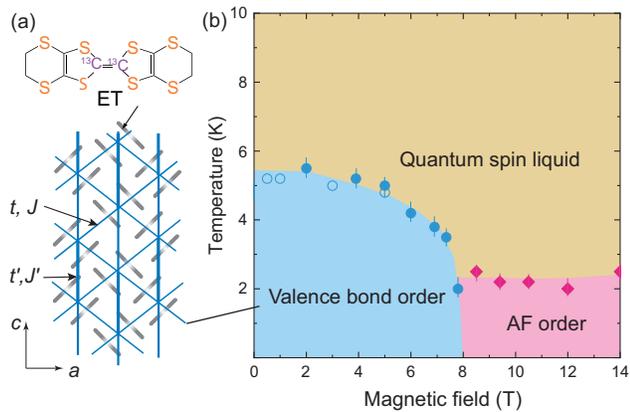}
	\caption{\label{Fig1} 
(a) Anisotropic triangular lattice of  molecular dimers in $\kappa$-(ET)$_2$B(CN)$_4$ with two kinds of transfers $t$ and $t^\prime$ above $T_c$. Spin-1/2 on each ET dimer is paired to form a valence bond crystal below $T_c = 5$ K. (b) Magnetic phase diagram based on NMR (closed symbols) and spin susceptibility $\chi$ (open symbols). Circles and diamonds with the uncertainty denote the valence bond order and antiferromagnetic transition temperatures, respectively, which are obtained from the temperature dependence of nuclear spin-lattice relaxation rate $T_1^{-1}$. 
	}
	\end{figure}

Molecular Mott insulators with a triangular lattice such as $\kappa$-(ET)$_2X$ and $X$[Pd(dmit)$_2$]$_2$ ($X$: counter ions) may provide the candidate showing the deconfined criticality. Depending on the strength and anisotropy of intermolecular interactions, a series of materials exhibit fertile ground states including quantum spin liquid \cite{Shimizu, Itou, Isono, Pratt, Hiramatsu, Kanoda}, antiferromagnetic order \cite{Miyagawa2, Shimizu2}, and valence bond order \cite{Yoshida, Tamura}. Some of them display superconductivity under pressure \cite{Shimizu3, Kurosaki}. Here we focus on a Mott insulator $\kappa$-(ET)$_2$B(CN)$_4$ having an anisotropic triangular lattice \cite{Yoshida}. The anisotropy of exchange interactions is evaluated as $J^\prime/J = 2$ ($J = 118$ K, $J^\prime = 236$ K) from the spin susceptibility \cite{Zheng}, where $J$ (or transfer integral $t$) and $J^\prime$ ($t^\prime$) form a square lattice and a chain, respectively [Fig. \ref{Fig1}(a)]. The anisotropy increases up to $J^\prime/J \sim 3$, as temperature ($T$) decreases. It agrees with the calculated transfer anisotropy $t^\prime/t = 1.4$ or $(t^\prime/t)^2$ = 2 \cite{Yoshida} and the optical measurement \cite{Mizukoshi}. Under pressure up to 2.5 GPa, the system remains Mott insulating. Thus the material is located far from the Mott transition at ambient pressure and provides an ideal Heisenberg antiferromagnet with anisotropic triangular lattice. 

The spin susceptibility of $\kappa$-(ET)$_2$B(CN)$_4$ exponentially decreases below $T_c \sim 5$ K without showing an indication of long-range magnetic order in the $^1$H NMR spectrum \cite{Yoshida}. Thus the ground state likely accompanies the valence bond order with translational symmetry breaking or the spin-Peierls transition, whereas the crystal structure has not been solved below $T_c$. The $^1$H NMR nuclear-lattice relaxation rate $T_1^{-1}$ exhibits weak $T$ dependence above $T_c$, consistent with the behavior of the low-dimensional quantum antiferromagnet in the quantum critical region \cite{Chubukov}. Considering the low $T_c$, an application of magnetic field is expected to induce quantum phase transition, which gives an insight into the origin of the valence bond order and the quasiparticle property. 

In this Article, we investigate the field-driven transition from the valence bond order phase in $\kappa$-(ET)$_2$B(CN)$_4$. The local field distribution and spin excitation are probed by $^{13}$C NMR spectroscopy down to low temperatures under magnetic field up to 14 T. Together with the magnetization and torque measurements, we uncover the ground state and low-energy properties across the field-driven transition. The magnetic field ($H$) versus $T$ phase diagram based on the experimental results [Fig. \ref{Fig1}(b)] is compared with the quantum spin liquid and spin-Peierls systems. 

\section{Experimental}

Single crystals of $\kappa$-(ET)$_2$B(CN)$_4$ were grown by an electrooxidation method \cite{Yoshida}. The magnetization was measured with a superconducting quantum interference device under 0.5 -- 7 T. The high-field magnetic torque for a single crystal was measured using a piezoresistive cantilever at the National High Magnetic Field Laboratory in Tallahassee. The magnetic field direction was tilted by 16$^\circ$ from the $ac$ plane. For NMR experiments, the 90\% $^{13}$C isotope was selectively enriched to the central double-bonded carbon sites of ET [Fig. \ref{Fig1}(a)] \cite{Larsen}. NMR measurements were conducted on a single crystal under the external magnetic field oriented along the $a$ axis ($<1^\circ$) of the conducting plane by utilizing a two-axis goniometer. The frequency-swept NMR spectrum was obtained from the spin echo measurement with the pulse width $t_{\pi/2} =$ 0.5 $\mu$s and $t_\pi = 1$ $\mu$s, and the sum was taken by 0.2 MHz steps covering the whole spectrum at low temperatures [Fig. 5(b)]. The nuclear magnetization recovery after the saturation followed a single exponential function above 5 K and becomes a stretched exponential function $M(t)/M_0 = 1-{\rm exp}[-(t/T_1)^\beta]$ with the exponent $\beta = 0.3- 1$ below 5 K, which is nearly independent of $H$. 

\section{Experimental Results}
\subsection{Spin susceptibility and Knight shift}

	\begin{figure}
	\includegraphics[scale=0.55]{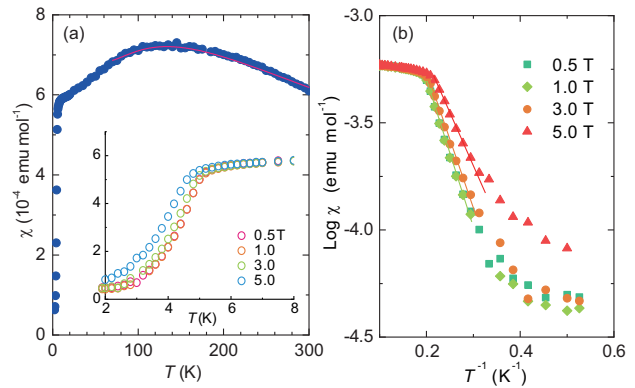}
	\caption{\label{Fig2} 
(a) Spin susceptibility $\chi$ for a polycrystalline sample of $\kappa$-(ET)$_2$B(CN)$_4$ at 1 T \cite{Yoshida} after subtracting the core diamagnetic susceptibility ($\chi_{\rm dia} = -4.52 \times 10^{-4}$ emu mol$^{-1}$). The red curve is a fitting result by a high-temperature series expansion \cite{Zheng}. Inset: Temperature dependence of $\chi$ at low temperatures under the magnetic field of 0.5, 1, 3, and 5 T. (b) Inverse temperature dependence of $\chi$ for the evaluation of spin gap with the exponential function $\chi \sim {\rm exp}(-\Delta/T)$. 
	}
	\end{figure}
The spin susceptibility $\chi$ of  $\kappa$-(ET)$_2$B(CN)$_4$ exhibits a broad maximum around 130 K \cite{Yoshida}, characteristic to the low-dimensional antiferromagnet, as shown in Fig. \ref{Fig2}(a). The $T$ dependence is fitted into a high-$T$ series expansion of the Heisenberg model with anisotropic triangular lattice ($J = 118$ K, $J^\prime = 236$ K) \cite{Zheng}. Below $T_c$ = 5 K,  $\chi$ measured at 1 T decreases continuously, indicating the second-order phase transition. The spin gap was evaluated as $\Delta$ = 27 K for the singlet-triplet model \cite{Yoshida} at 1 T. Here we employ the simple exponential function [Fig. \ref{Fig2}(b)], yielding $\Delta$ = 16, 14.8, and 11.1 K under $H$ = 1, 3, and 5 T, respectively. The negligible Curie tail increase in $\chi$ at low temperatures certificates the high crystal quality.

	\begin{figure}
	\includegraphics[scale=0.52]{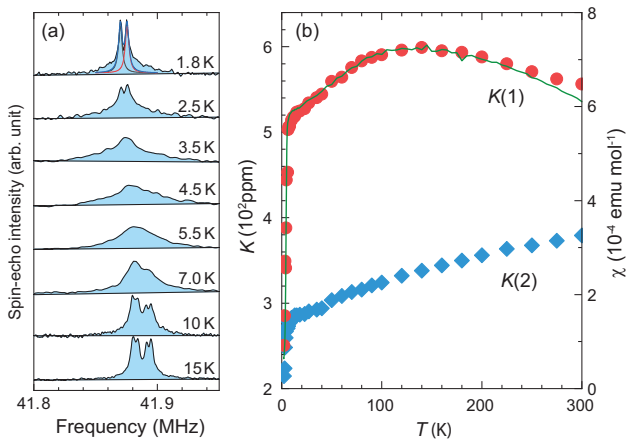}
	\caption{\label{Fig3} 
(a) $^{13}$C NMR spectra at $H$ = 3.9 T along the $a$ axis in $\kappa$-(ET)$_2$B(CN)$_4$. Lorentzian fitting curves are shown at the lowest temperature. (b) Knight shifts, $K(1)$ and $K(2)$, for two carbon sites, C(1) and C(2), which are measured from the reference (tetramethylsilane). A green curve denotes the spin susceptibility $\chi$ at 1.0 T (the right-hand axis) \cite{Yoshida}. 
	}
	\end{figure}	

	The $^{13}$C NMR spectrum provides the local spin susceptibility and the local field distribution at nuclear sites through the hyperfine interaction. In $\kappa$-(ET)$_2$B(CN)$_4$, each ET molecule has two $^{13}$C sites, all of which in the unit cell (orthorhombic $Pnma$) are equivalent [Fig. \ref{Fig1}(a)]. Therefore, the NMR spectrum in Fig. \ref{Fig3}(a) consists of two doublets due to dipole-coupled two $^{13}$C spins \cite{Abragam}. Upon cooling, the spectrum gradually broadens toward $T_c$, coinciding with the intensity reduction due to fluctuations and the inhomogneous field due to the generation of paramagnetic domains around the structural transition. Below $T_c$, the spectrum turns to be narrow and remains only the doublet splitting. The splitting frequency is about $3/2$ times larger than that observed at high temperatures, which represents a crossover from the unlike-spin to like-spin coupling, as the spin susceptibility vanishes below $T_c$ \cite{Abragam}. Since $^{13}$C spins ($I =1/2$) do not have electric quadrupole interaction, structural fluctuations is observed through a small nuclear dipole coupling and paramagnetic domains. 

	The residual broad tails of the spectrum at low temperatures can be attributed to staggered moments nucleated around the domains through the Dzyaloshinkii-Moriya (DM) interaction \cite{Shimizu2}. The sharp central line means that the majority of spins remain nonmagnetic. 
	
The Knight shift $K$ proportional to the local spin susceptibility $\chi_{\rm loc}$ is defined for two carbon sites on each ET molecule, termed as $K(1)$ and $K(2)$. As shown in Fig. \ref{Fig3}(b), $K(1)$ exhibits a broad maximum around 130 K as observed in $\chi$, reflecting the short-range antiferromagnetic correlation. In contrast, $K(2)$ monotonously decreases upon cooling. The site dependence implies that the hyperfine coupling and the chemical shift vary with $T$ due to the anisotropic lattice contraction \cite{Yoshida}. 

	\begin{figure}
	\includegraphics[scale=0.55]{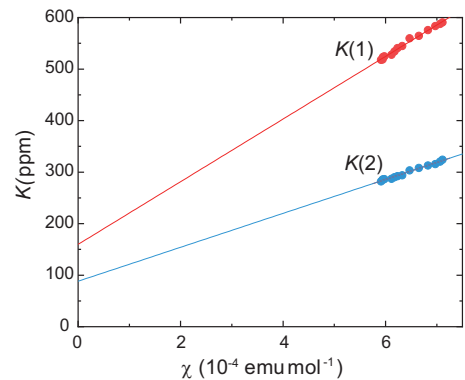}
	\caption{\label{Fig4} 
$^{13}$C Knight shifts plotted against $\chi$ for 10 -- 50 K. The linearity yield the hyperfine coupling constant for two $^{13}$C sites. The $y$-intercept gives the chemical shift. 
	}
	\end{figure}	

At low temperatures below 50 K, the Knight shifts well scale to $\chi$. The linearity in the $K-\chi$ plot for $10-50$ K yields the hyperfine coupling constant $H_{\rm hf}$, as shown in Fig. \ref{Fig4}. In the present field direction ($H \parallel a$), we obtained  $H_{\rm hf}$ = 0.34(2) T/$\mu_{\rm B}$ for C(1) and 0.18(4) T/$\mu_{\rm B}$ for C(2). The difference reflects the asymmetric spin density distribution due to the molecular dimerization \cite{Shimizu2}. Here $H_{\rm hf}$ is composed of the Fermi contact and dipolar interaction terms. The anisotropic dipole term may depend on temperature when the principal axis governed by the strength of the dimerization changes upon cooling. Similar behavior has been observed in the other members of $\kappa$-(ET)$_2$X \cite{Shimizu2, Shimizu4}.

\subsection{Magnetic field driven transition}

	\begin{figure}
	\includegraphics[scale=0.8]{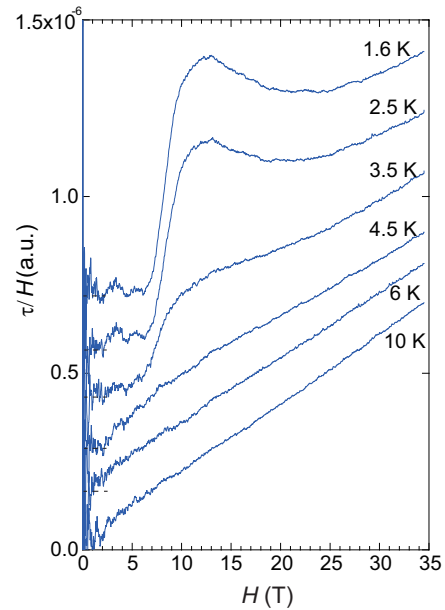}
	\caption{\label{Fig5} 
Magnetic torque $\tau$ divided by magnetic field $H$ at low temperatures. The data at each temperature include a constant offset for clarity. 
	}
	\end{figure}

The field sensitivity of the spin gap implies that a quantum phase transition occurs at the critical field $H_c$ where $\Delta$ goes to zero or competes with the other order parameter. For investigating the field-driven transition, we measured the magnetic torque $\tau$ responsive to the transverse magnetization, $\tau/H \sim M_T$, as shown in Fig. \ref{Fig5}. In a paramagnetic state above $T_c$, $\tau/H$ is governed by the $g$-value anisotropy and increases with the external field $H$. The slop is nearly independent of temperature at high fields. 

Below 4.5 K, $\tau/H$ becomes nonlinear to $H$. $\tau/H$ behaves constantly at low fields, reflecting $\chi \simeq 0$ with a negligible Curie impurity component. Then $\tau/H$ starts to increase around 8 T, which clearly indicates a magnetic transition from the nonmagnetic state into the magnetically ordered state or the paramagnetic spin liquid state. Below 2.5 K, a maximum appears in $\tau/H$ around 11--12 T. After showing a minimum around 24--28 T, $\tau/H$ linearly increases with $H$ without saturation up to 35 T. However, the nature of the strong field-induced phase is unclear at the present stage.

	\begin{figure}
	\includegraphics[scale=0.7]{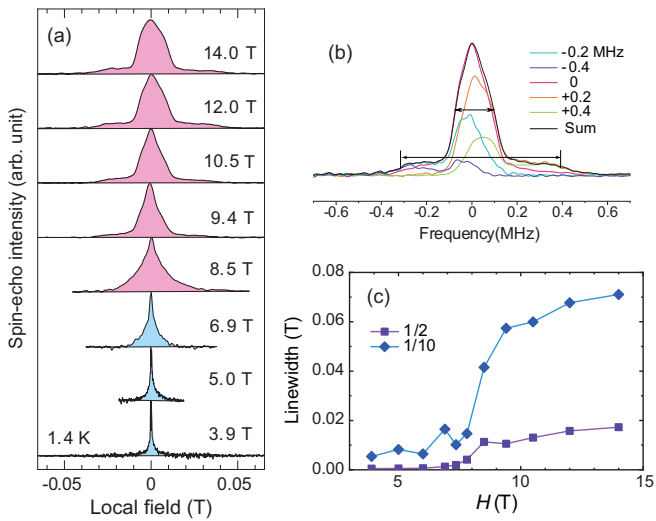}
	\caption{\label{Fig6} 
	(a) Magnetic field dependence of the $^{13}$C NMR spectrum for $\kappa$-(ET)$_2$B(CN)$_4$ at 1.4 K. Frequency $\omega$ was converted to the local field $\Delta B$ using a relation $\Delta B = \Delta \omega/\gamma_0$ ($\gamma_0 = 10.7054$ MHz/T). (b) Definition of the 1/2 and 1/10 linewidths for $T$ = 1.4 K and $H$ = 12 T, where the spectrum is constructed as a sum of Fourier transformed spin-each intensities taken with a 0.2 MHz step. (c) Magnetic field dependence of the 1/2 and 1/10 linewidths. 
	}
	\end{figure}
	
	The magnetic field dependence of $^{13}$C NMR spectrum was measured under the magnetic field of 3.9 -- 14 T at 1.4 K, as displayed in Fig. \ref{Fig6}(a). The spectrum gradually spreads even at low fields, indicating the evolution of the local field distribution upon increasing $H$. Above 8.5 T, the spectral broadening becomes prominent at low temperatures. The NMR spectrum eventually changes into a two-step bell shape. The emergence of the huge local field signals long-range magnetic order. Under $H$ = 14 T, the maximum local field reaches $\pm 0.04$ T at the tails of the spectrum. 
	
	As shown in Fig. \ref{Fig6}(b,c), the linewidths defined at the 1/2 and 1/10 of the maximum intensity evolve above the critical field $H_c \sim $ 8 T and levels off at high fields, consistent with the magnetic torque result. The 1/2 and 1/10 widths measure the evolution of the order parameter at two $^{13}$C sites with the different $H_{\rm hf}$. Indeed, the 1/2 width (0.03 T) and 1/10 width (0.07 T) approximately scale to $H_{\rm hf}$ of two $^{13}$C sites. The result yields the maximum moment of $\sim$0.1 $\mu_{\rm B}$, which continuously grows above $H_c$ and saturates at high fields.

	\begin{figure}
	\includegraphics[scale=0.4]{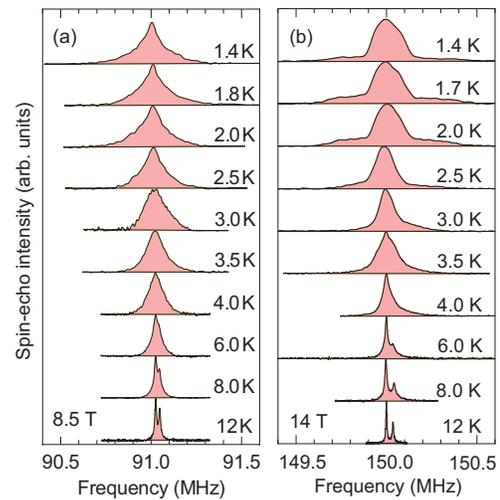}
	\caption{\label{Fig7} 
Temperature dependence of the $^{13}$C NMR spectrum for the single crystal of $\kappa$-(ET)$_2$B(CN)$_4$ under the magnetic field of (a) 8.5 and (b) 14 T along the $a$ axis.   %
	}
	\end{figure}

		The temperature dependence of NMR spectra above $H_c$ is shown in Fig. \ref{Fig7}. At 8.5 T, the spectral broadening starts below 10 K and continues upon cooling, while leaving a narrow line at the central position [Fig. \ref{Fig7}(a)]. It suggests the coexistence of the magnetic phase and the valence bond order phase, or the formation of soliton lattice as observed in the 1D chains \cite{Horvatic}. However, the singular edge peak characteristic to solitons was not detected within the experimental uncertainty. At 14 T, the site-dependent broadening becomes clearer above $T_{\rm N}$. The asymmetric spectral shape in the paramagnetic state becomes nearly symmetric below 2.5 K, which gives evidence for the magnetic order transition.  

	In reference to the spin structure of $\kappa$-(ET)$_2$Cu[N(CN)$_2$]Cl belonging to the same space group as $\kappa$-(ET)$_2$B(CN)$_4$, the colinear magnetic order splits the $^{13}$C NMR spectrum under the magnetic field along the easy axis ($H \parallel c$) and gives only a positive shift for the perpendicular direction ($H || a$) \cite{DM}. The observed continuous and symmetric broadening points to an incommensurate magnetic structure such as a spiral order expected for a triangular lattice antiferromagnet \cite{Tocchio,Mizusaki,Ghorbani}. The full determination of the spin structure requires the $H_{\rm hf}$ tensor and the angular dependence measurements. The sizable moment contraction highlights a crucial role of quantum fluctuations for the system nearby the disordered state. 

\subsection{Spin dynamics across field driven transition}

	\begin{figure}
	\includegraphics[scale=0.6]{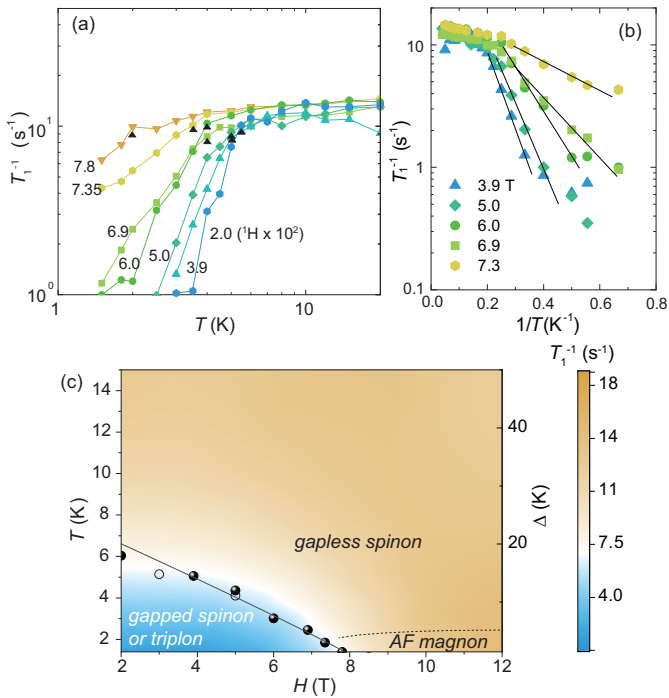}
	\caption{\label{Fig8} 
(a) Temperature dependence of the $^{13}$C nuclear spin-lattice relaxation rate $T_1^{-1}$ at 3.9--7.8 T. $^1$H NMR $T_1^{-1}$ (a long component) at 2.0 T scale to $^{13}$C data by multiplying $10^2$ \cite{Yoshida}. $T_c$ defined by the onset of $T_1^{-1}$ drop is marked by arrows. (b) Inverse temperature dependence of $T_1^{-1}$ for evaluating the spin gap $\Delta$. (c) Field dependence of $\Delta$ (the right-hand axis) and the contour plot of $T_1^{-1}$ obtained from the temperature and field dependence measurement. Solid and open circles are evaluated from $T_1^{-1}$ and $\chi$, respectively. }
	\end{figure}

The spin excitation is investigated by the nuclear spin-lattice relaxation rate $T_1^{-1}$ across $H_c$. As shown in Fig. \ref{Fig8}, $T_1^{-1}$ is nearly independent of $T$ above $T_c$, as expected in low-dimensional quantum antiferromagnets for a $T$ range lower than the exchange coupling ($T < J$) \cite{Chubukov}. The behavior differs from that observed in a triangular lattice system having the spin liquid ground state, as discussed below. $T_1^{-1}$ decreases steeply as the spin gap $\Delta$ opens below $T_c$ in a low field range below 8 T. $T_1^{-1}$ is enhanced with increasing $H$, as $T_c$ and $\Delta$ are suppressed toward $H_c \sim 8$ T. The obtained field dependence of $T_c$ is plotted in Fig. \ref{Fig1}(b), consistent with the result of $\chi$. 

We evaluated $\Delta$ by fitting $T_1^{-1}$ into an exponential function $\sim$ exp$(-\Delta/T)$ [Fig. \ref{Fig8}(b)]. We obtained $\Delta = 14.5, 12.0, 7.2 $ K at $H$ = 3.9, 5.0, 6.9 T, respectively, and plotted in Fig. \ref{Fig8}(c). The results are consistent with those obtained from $\chi(T)$ [Fig. \ref{Fig2}(b)]. Since $T_1^{-1}$ measures the dynamical spin correlation summed over the wave vector space, the result represents the presence of the spin excitation gap independent of the wave vector. We find that $\Delta$ linearly decreases with increasing $H$, $\Delta \sim (H_c - H)^z$, with $H_c \sim 8.2(2)$ T and the critical exponent $z = 1.0(3)$. The $H$ and $T$ dependence of $T_1^{-1}$ is also illustrated as a contour plot [Fig. \ref{Fig4}(c)]. Approaching $H_c$, the $T$ independent region persists to low temperatures, as expected in the gapless spinon regime. 

	\begin{figure}
	\includegraphics[scale=0.38]{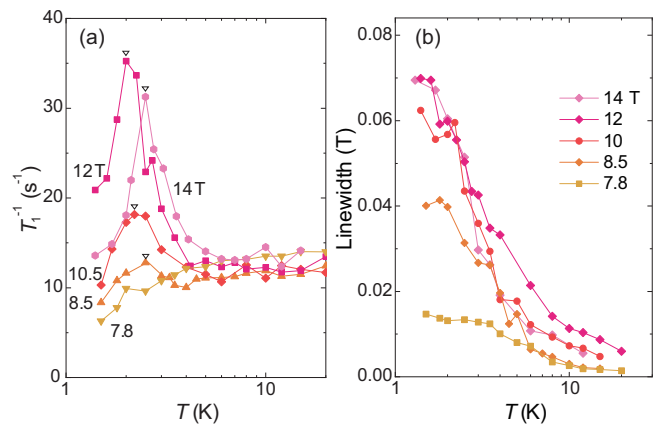}
	\caption{\label{Fig9} 
(a) Temperature dependence of $T_1^{-1}$ above 7.8 T. Arrows point to the magnetic order transition at $T_{\rm N}$. (b) Linewidth of $^{13}$C NMR spectrum, defined by 1/10 of the peak intensity in Fig. \ref{Fig6}(b). 
	}
	\end{figure}

Above $H_c$, $T_1^{-1}$ exhibits a peak structure at $2 - 2.5$ K due to the slowing-down of spin fluctuations, as shown in Fig. \ref{Fig9}(a). The peak determines the magnetic order transition temperature $T_{\rm N}$. It is nearly independent of $H$, as plotted in Fig. \ref{Fig1}(b). Below $T_{\rm N}$ one can expect an increase in the linewidth of the $^{13}$C NMR spectrum, which measures the evolution of the order parameter. As shown in Fig. \ref{Fig9}(b), the linewidth remains small at 7.8 T just below $H_c$. The spectral broadening starts from $T > T_{\rm N}$ above 8.5 T due to fluctuations and inhomogeneity, while the broadening levels off below $T_{\rm N}$ defined by $T_1^{-1}$.

\section{Discussion}
In this section, we focus on the nature of the dynamical spin correlation in the paramagnetic spin liquid phase. We also discuss the origin for the valence bond order transition based on the magnetic phase diagram as well as the possible magnetic phase at high fields. 

\begin{figure}
	\includegraphics[scale=0.33]{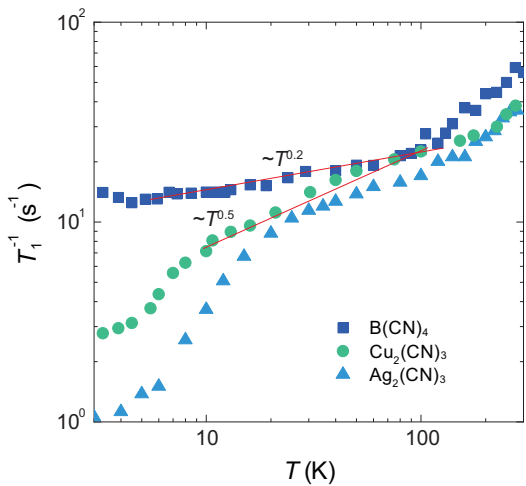}
	\caption{\label{Fig10} 
Temperature dependence of $T_1^{-1}$ for $^{13}$C NMR in $\kappa$-(ET)$_2X$, $X$ = B(CN)$_4$, Cu$_2$(CN)$_3$ \cite{Shimizu4}, and Ag$_2$(CN)$_3$ \cite{Shimizu2}. 
	}
	\end{figure}

The series of $\kappa$-(ET)$_2X$ [$X$ = Cu$_2$(CN)$_3$, Ag$_2$(CN)$_3$] have been extensively investigated as frustrated quantum antiferromagnets with triangular lattice close to Mott transition \cite{Shimizu,Shimizu2,Kanoda}, where the anisotropy $t^\prime/t$ is close to or smaller than unity. $\kappa$-(ET)$_2$B(CN)$_4$ is the rare example of the frustrated triangular lattice with one-dimensional anisotropy ($t^\prime/t > 1$). Thus it is interesting to compare the difference in the dynamical spin correlation between these systems. In the temperature range lower than the antiferromagnetic exchange interaction, the system is regarded as quantum spin liquid involving low-lying excitations. $T_1^{-1}$ measures the dynamical spin susceptibility in the low energy limit \cite{Chubukov, Senthil2}. On the isotropic triangular lattice such as $\kappa$-(ET)$_2$Ag$_2$(CN)$_3$ showing a quantum spin liquid state, $T_1^{-1}$ follows a power law $\sim T^\eta$ ($20 < T < 200$ K) with the anomalous exponent $\eta = 0.4-0.5$, as shown in Fig. \ref{Fig10} \cite{Shimizu2,Shimizu4}. Interestingly, the exponent is close to that expected for deconfined spinon excitations ($\eta = 0.6$) in the frustrated model \cite{Senthil2}.

    In contrast, $T_1^{-1}$ for $\kappa$-(ET)$_2$B(CN)$_4$ follows a power law $\sim T^\eta$ with the small exponent $\eta = 0.19(1)$ in a wide range ($5 < T < 100$ K), as shown in Fig. \ref{Fig10}. Here the strong frustration effect is manifested in the low $T_c$ and $T_{\rm N}$ compared with the exchange coupling. The exponent is rather close to that of the 1D quantum critical antiferromagnet, where $T_1^{-1}$ behaves $T$-independent ($\sim T^\eta$, $\eta = 0$), as the antiferromagnetic correlation length $\xi$ diverges with $\xi \sim T^{-z}$ (the dynamical exponent $z = 1$) at the specific wave vector \cite{Chubukov}. The behavior was observed in Sr$_2$CuO$_3$ \cite{Takigawa2} and (TMTTF)$_2$PF$_6$ \cite{Brown}. 

On the magnetic phase diagram [Fig.\ref{Fig1}(b)], the valence bond order phase is suppressed toward the critical field $H_c \sim 8$ T, as observed by magnetic torque and NMR spectrum measurements. It is also consistent with the field dependence of the spin gap $\Delta$ obtained from $T_1^{-1}$. The continuous suppression of $\Delta$ implies the quantum phase transition in the frustrated spin system \cite{Senthil, Scammell, Jian}. However, the incommensurate magnetic order phase appears from the finite temperature across $H_c$. It implies that the second order valence bond order transition is terminated by the first-order magnetic transition at low temperatures as anticipated from the mean-field theory \cite{Cross, Inagaki}. 

 The shape of phase diagram is similar to those of the quasi-1D spin-Peierls system such as CuGeO$_3$ and (TMTTF)$_2$PF$_6$\cite{Hase, Horvatic, Brown, Brown2}. It suggests that the valence bond order occurs through the lattice dimerization along the $c$ axis. Indeed, the crystal structure becomes more 1D upon cooling \cite{Yoshida}. This tendency is distinct from that observed in the other $\kappa$-(ET)$_2X$ members with $t^\prime/t <1$ \cite{Kanoda}. Thus the renormalized 1D spin correlation can develop in the gapless spin liquid regime above $T_c$ \cite{Ogata} and eventually couple the lattice degrees of freedom. Furthermore, the discrete (not polymeric) counter ion and the absence of the structural disorder allows the spin dimerization along one direction.

Focusing on the magnetically ordered phase at high fields, the magnetic structure would be highly degenerate and nontrivial on the triangular lattice. Theoretical calculations have been reported based on the Hubbard and Heisenberg model as a function of anisotropy of triangular lattice, $t^{\prime}/t$ or $J^{\prime}/J$ \cite{Kaneko, Tocchio,Ghorbani}. For $J^{\prime}/J > 1$, the magnetic ground state of Mott insulator can be both quantum spin liquid and antiferromagnetic orders such as colinear and spiral spin structures. Indeed, the isostructure Mott insulator $\kappa$-(ET)$_2$CF$_3$SO$_3$ ($t^{\prime}/t = 1.8$) exhibits the magnetically ordered ground state with an incommensurate structure below $T_{\rm N} = 2.5$ K \cite{Ito}. Therefore, the present system is naturally expected to exhibit the long-range magnetic order, as the valence bond crystal is suppressed by the magnetic field.  

\section{Conclusion}
In conclusion, the field-induced quantum phase transition was observed from the valence bond order to the antiferromagnetic long-range order state by $^{13}$C NMR and magnetization measurements in the Mott insulator $\kappa$-(ET)$_2$B(CN)$_4$ having a distorted triangular lattice. We found that the spin excitation gap is linearly suppressed by magnetic field toward the critical point. As the valence bond order phase melts, the quantum spin liquid with gapless excitation appears above the critical field. The obtained phase diagram manifests the competing magnetic ground states in the frustrated magnet. The significant role of quantum fluctuations are highlighted as the reduction of the antiferromagnetic moment and the persistent low-lying excitation down to low temperatures.  
	
\section*{Acknowledgements}
	This work was financially supported by Grants-in-Aid for Scientific Research (Grants No. JP16H02206, JP16H04012, JP17H05151, JP18H04223, JP19H05824, JP19H01837, and JP20H02709) from JSPS, and partly supported by the Kyoto University Foundation. 

\bibliography{Bib_BCN}% Produces the bibliography via BibTeX.

\end{document}